\documentclass[11pt]{amsart}
\usepackage{geometry}                
\usepackage{cite}

\usepackage{amsfonts,amssymb,amsmath,calc,graphicx}

\usepackage{bytefield}
\usepackage{setspace}
\usepackage{color}
\newenvironment{nouppercase}{%
  \renewcommand{\uppercasenonmath}[1]{}}{}

\def\m#1{\mathbf{#1}}
\newcommand{\tab}{\hspace*{1.5em}}

\newtheorem{theorem}{Theorem}
\newtheorem{lemma}{Lemma}

\begin{document}

\title{Proportional Fair MU-MIMO in 802.11 WLANs}

\author{V. Valls and D. J. Leith\\Hamilton Institute, NUI Maynooth}
\thanks{This work appeared in \textit{IEEE Wireless Communications Letters}, vol.3, no.2, pp.221-224, April 2014. doi: 10.1109/WCL.2014.020314.130884}
\thanks{This material is based upon works supported by the Science Foundation Ireland under Grant No. 11/PI/1177.}
\begin{nouppercase}
\maketitle
\end{nouppercase}

\begin{abstract}
We consider the proportional fair rate allocation in an 802.11 WLAN that supports multi-user MIMO (MU-MIMO) transmission by one or more stations.   We characterise, for the first time, the proportional fair allocation of MU-MIMO spatial streams and station transmission opportunities.   While a number of features carry over from the case without MU-MIMO, in general neither flows nor stations need to be allocated equal airtime when MU-MIMO is available.
\end{abstract}


\section{Introduction}
 
The next generation of 802.11 WLANs are expected to support MU-MIMO transmission, whereby parallel transmissions can be simultaneously made to multiple stations.   This significantly extends the MIMO support introduced by the 802.11n standard and is, for example, included as part of the current draft 802.11ac standard that aims to support wireless data rates in excess of 1 Gbps.    MU-MIMO offers much greater flexibility in scheduling MIMO transmissions, but immediately raises the question of how best to allocate MIMO spatial streams amongst network flows so as to balance fairness and performance.  In this paper we consider the proportional fair allocation in an 802.11 WLAN that supports MU-MIMO transmission by one or more stations.    While proportional fairness \cite{Kelly:1998:0160-5682:237} in WLANs has been the subject of considerable interest in the literature, it has only recently been put on a rigorous basis in \cite{5910091} and consideration of MU-MIMO is new.

The main contribution of the paper is to rigorously characterise, for the first time, the proportional fair allocation of spatial streams and station transmission opportunities in WLANs where one or more stations support MU-MIMO.   We demonstrate that this allocation can be found in a distributed manner, without the need for message passing.   We show that a number of features carry over from the case without MU-MIMO, specifically that the rate region boundary is characterised by the station total airtimes summing to unity and that stations carring the same number of flows then stations are assigned equal total airtime. Importantly, however, we find that MU-MIMO generally leads to a qualitatively different allocation of airtime compared to the situation without MU-MIMO.  Namely, in general flows are \emph{not} allocated equal airtime.   This is because it is the station total airtime that corresponds to the shared network resource being consumed and so to the ``cost'' of transmissions.  When MU-MIMO transmission is available, flow transmissions occur in parallel and so multiple flows can share the same station airtime.

\section{Related Work}
In \cite{cai2008distributed} the authors propose a novel MAC design with opportunistic MU-MIMO scheduling based on channel sounding feedback, where packets are selected depending on their transmission duration and type of traffic. In \cite{5684351} is also proposed a novel MAC design for MU-MIMO that focusses on issues such as MAC ACKing of MU-MIMO transmissions. Packets are scheduled for transmission using a weighted queuing mechanism that considers both packets acknowledgements and type of traffic. However, in both \cite{cai2008distributed} \cite{5684351} fairness and allocation of MU-MIMO transmission patterns amongst flows is not considered. The work in \cite{6287486} focusses on packet aggregation in an IEEE 802.11ac AP, and considers a  fixed MU-MIMO schedule where one flow is allocated per spatial stream. Regarding utility fairness in WLANs, in \cite{5910091} is presented the first rigorous analysis of proportional fairness in 802.11 WLANs where transmissions are to a single destination. 
 
\section{Network Model}
\subsection{Preliminaries}
We take as our starting point the network model in \cite{5910091}. Consider an 802.11 WLAN with $n$ stations, where each station $i$ attempts to transmit at each MAC slot with probability $\tau_i$. We will assume that stations are configured with $CW_{min} = CW_{max}$ (which is supported in 802.11 starting with 802.11e/WME in 2005, 802.11-2007 and subsequent standards), so that the attempt probability is independent of the success or failure of the last transmission. Moreover, it is also assumed that there are no hidden terminals, so all nodes in the network can sense any ongoing transmission.  Because of this, a collision can only happen if two or more stations transmit in the same slot. We also assume that noise-induced losses are negligible, although this assumption could be relaxed.

The probability that a transmission by station $i$ is successful is the probability that only station $i$ transmits and is given by $P_{succ,i} =  \tau_i  \prod_{k=1,k \ne i}^n (1-\tau_k)$. The probability that a MAC slot is idle is given by the probability that none of the stations transmit, $P_{idle} = \prod_{k=1}^n (1- \tau_k)$. Finally, the probability that a transmission by station $i$ collides is $P_{coll,i} = 1- P_{succ,i} - P_{idle}$. The throughput of station $i$ is then given by
\begin{align}\label{eq:model}
S_i(\boldsymbol \tau) 
= & \frac{P_{succ,i}D_i}{\sigma P_{idle} + T_s (1-P_{idle})},
\end{align}
where $\sigma$ is the duration of an idle slot, $T_s$ the duration of a busy slot (either successful or collision) and $D_i$ is the size in bits of the frame payload of station $i$.   Throughput model (\ref{eq:model}) is standard, see e.g.  \cite{5910091,6009216,4100720} and references therein.

Let $x_i= \tau_i / (1-\tau_i)$, thus $x_i \in [0,\infty)$ as $\tau_i \in [0,1]$. With this change of variable we have that $P_{idle} = 1 / \prod_{k=1}^n (1+x_k) $,  $P_{succ,i} = x_i P_{idle}$, and 
\begin{align}
S_i(\m{x}) = & \frac{x_i}{X(\m{x})}\frac{D_i}{T_s},  \label{eq:throughput}
\end{align} where $X(\m{x}) = a + \prod_{k=1}^n (1+x_k)-1$ with $a =  \sigma / T_s$ and $\m{x} = \left[ x_1, \dots, x_n \right]$. Notice that $x_i / X(\m{x})$ is the successful airtime for station $i$, and $D_i / T_s $ the rate. Hence, the total airtime  ($T_i$) of station $i$ is given by the airtime spent on successful transmissions and collisions
\begin{align}
T_i  
=  & \frac{x_i}{X(\m{x})}\left( 1+ \frac{P_{coll,i}}{1-P_{coll,i}} \right) \label{eq:airtime}.
\end{align}

\subsection{Extension to MU-MIMO}
\label{sec:muextension}

The throughput model (\ref{eq:model}) can be extended as follows to encompass MU-MIMO, where stations can transmit multiple spatial streams simultaneously. Let $F_i$ be the set of flows carried by station $i$, and $F = \cup_{i=1}^n F_i$  the set of flows in the WLAN.  We let vector $\m{v}_{ik}$ describe the $k^{th}$ MU-MIMO transmission pattern on station $i$, where $\m{v}_{ik}$ has $|F_i|$ elements, and element $v_{ikf}$ defines the number of spatial streams allocated to flow $f$ in this pattern. We collect the set of $K_i$ possible transmission patterns for station $i$ together to form matrix $\m{V}_i$, where the $k^{th}$ row of $\m{V}_i$ describes the $k^{th}$ pattern, $k=1,...,K_i$. See for example Figure \ref{fig:combinations}.     The set of allowable transmission patterns will be determined by the network characteristics, i.e. number of antennas of the stations,  channel conditions and protocol constraints.  For example, the draft 802.11ac restricts the use of MU-MIMO to the AP and allows at most 8 spatial streams with at most 4 streams for one client station. However, to keep our analysis as general as possible we will not make any assumptions about the structure of matrix $\m{V}$.
\begin{figure}[h!]
\centering
\begin{bytefield}[bitheight=1\baselineskip ]{127}\selectfont \small
\bitbox[]{5}{} \bitbox[]{4}{$f_1$} & \bitbox[]{4}{$f_2$} & \bitbox[]{4}{$f_3$} & \bitbox[]{4}{$f_4$} 
\end{bytefield}\\
\begin{bytefield}[bitheight=1.1\baselineskip ]{127} \fontfamily{phv}\selectfont \small
\bitbox[]{5}{$k=1$} \bitbox{4}{2} & \bitbox{4}{0} & \bitbox{4}{1} & \bitbox{4}{1} 
\end{bytefield}\\
\begin{bytefield}[bitheight=1.1\baselineskip ]{127} \fontfamily{phv}\selectfont \small
\bitbox[]{5}{$k=2$} \bitbox{4}{1} & \bitbox{4}{2} & \bitbox{4}{1} & \bitbox{4}{0} 
\end{bytefield}\\ \setstretch{0.25}
\begin{bytefield}[bitheight=3.5\baselineskip]{127} \fontfamily{phv}\selectfont \small
\bitbox[]{5}{} \bitbox[]{4}{.\\.\\.} & \bitbox[]{4}{.\\.\\.} & \bitbox[]{4}{.\\.\\.} & \bitbox[]{4}{.\\.\\.} 
\end{bytefield}\\ \setstretch{1.1}
\begin{bytefield}[bitheight=1\baselineskip ]{127} \fontfamily{phv}\selectfont \small
\bitbox[]{5}{$k=K_i$} \bitbox{4}{1} & \bitbox{4}{1} & \bitbox{4}{1} & \bitbox{4}{1} 
\end{bytefield}
\caption{Example of MU-MIMO transmission matrix $\m{V}_i$ where each row represents a possible MU-MIMO transmission pattern for station $i$. For example, row 1 defines a pattern where two spatial streams are allocated to flow $f_1$, and one spatial stream each to flows $f_3$ and $f_4$.}
\label{fig:combinations}
\end{figure}

Next, let $\pi_{ik}$ denote the fraction of transmission opportunities that pattern $k$ is selected by station $i$, with $\sum_{k=1}^{K_i}\pi_{ik}=1$.  We collect the $\pi_{ik}$ for station $i$ together in vector $\boldsymbol \pi_i$. We can then express the throughput of flow $f$ on station $i$ as 
\begin{align}
s(f) = \frac{x_i}{X(\m{x})}\frac{D_f}{T_s} \sum_{k=1}^{K_i} \pi_{ik}  v_{ikf}   \label{eq:flownonconvex},
\end{align}
where $\sum_{k=1}^{K_i} \pi_{ik}  v_{ikf}$ is the average number of spatial streams used by flow $f$ in station $i$, $D_f \sum_{k=1}^{K_i} \pi_{ik}  v_{ikf}$ is the average number of bits sent for a flow $f$ in a successful transmission and $D_f$ is the number of bits transmitted by flow $f$ on a single spatial stream in a successful transmission. Note that since spatial streams are transmitted in parallel, a MU-MIMO transmission occupies the same amount of airtime as a single spatial stream and so the total airtime $T_i$ used by station $i$ is still given by (\ref{eq:airtime}).

In (\ref{eq:flownonconvex}) the number of bits transmitted ($D_f$) by flow $f\in F_i, \ i=1,\dots,n$ does not depend on the selected transmission pattern. However,  due to varying channel conditions (such as inter-user interference) it is  likely that the number of bits transmitted by flow $f$ varies with the transmission pattern used.  Our model can be easily extended to include this.  Let matrix $\m{D}_i \in \mathbb R_+^{K_i \times |F_i|}, \ i=1,\dots,n$ denote the average number of bits of flow $f \in F_i$ in a transmission pattern $k=1,\dots,K_i$. Then, the throughput expression of a flow $f$ is given by
\begin{align}
s(f) = \frac{x_i}{X(\m{x})}\frac{1}{T_s} \sum_{k=1}^{K_i} \pi_{ik}  v_{ikf} d_{ikf}, \label{eq:flownonconvex2}
\end{align}
Nevertheless, as this generalisation is straightforward, we use (\ref{eq:flownonconvex}) rather than (\ref{eq:flownonconvex2}) for the rest of the analysis to streamline notation.

\section{Proportional Fair Rate Allocation}
\subsection{Log-convexity}

It can be readily verified that the flow throughput (\ref{eq:flownonconvex}) is non-convex in $\m{x}$, and also in $\m{\tau}$.   Fortunately, however, we have the following:
\begin{lemma}[Log-convexity]\label{lem:convexity}
\begin{align}
-\tilde{x}_i - \log \left( \sum_{k=1}^{K_i} \pi_{ik} v_{ikf} \right) - \log \frac{D_f}{T_s} + \log X(e^{\tilde{\m{x}}}) \label{eq:convexity}
\end{align}
is convex in $\tilde{\m{x}}$ and $\boldsymbol \pi$, where $\tilde{\m{x}}=[\tilde{x}_1,\dots,\tilde{x}_n]^T$, $\tilde{x_i} = \log x_i$.
\begin{proof}
Observe that the first term is linear in $\tilde{\m{x}}$ (and so convex), the second term is convex in $\boldsymbol \pi$ due to the convexity of the negative log function when composed with a linear map \cite{boyd2004convex}. The last term is convex in $\tilde{\m{x}}$  by Lemma 1 of \cite{5910091}. 
\end{proof}
\end{lemma}

\subsection{Utility-fair optimisation}
The proportional fair rate allocation is the solution to the utility-fair optimisation problem $P$:
\begin{align}
\underset{\tilde{\m{x}},\tilde{\m{s}},\boldsymbol \pi}\max \tab & \sum_{f \in F} \tilde{s}(f)  \label{eq:objective} \\
\text{s.t.\tab} & {\tilde{s}(f)} \le \log \Bigg (\frac{e^{\tilde{x}_i}}{X(e^{\tilde{\m{x}}})}\frac{D_f}{T_s} \sum_{k=1}^{K_i} \pi_{ik} v_{ikf}  \Bigg ),\ f\in F_i  \label{eq:c1}\\
& \sum_{k=1}^{K_i} \pi_{ik} = 1, \tab \pi_{ik} \ge 0,\ i=1,\dots,n \label{eq:c3} 
\end{align}
where $\tilde{s}(f) = \log s(f)$, $\tilde{x_i} = \log x_i$.    

It follows from Lemma \ref{lem:convexity} that constraint (\ref{eq:c1}) is convex.   Since the objective and remaining constraints are linear in the transformed variables, the optimisation problem is convex and so a proportional fair allocation exists.    The proportional fair rate allocation is almost completely characterised as follows: 
\begin{theorem}[Proportional Fairness]\label{lem:kkt}
The MU-MIMO proportional fair rate allocation is characterised by: (i) the airtime allocated to station $i$ is $T_i= \frac{ |F_i|}{ |F|}$ where $|F_i|$ is the number of flows carried by station $i$ and $|F|$ the total number of flows in the WLAN, (ii)  the station total airtimes sum to unity $\sum_{i=1}^n T_i=1$, (iii) the allocation of MU-MIMO transmission patterns on station $j$ satisfies 
\begin{align}
\sum_{f \in F_j} \lambda_f \frac{v_{jlf} }{\sum_{k=1}^{K_j} \pi_{jk} v_{jkf}} =  \nu_j - \theta_{jl},\ l=1,\dots,K_j \label{eq:3kkt}
\end{align}
where $\nu_j$, $\theta_{jl}$ $j=1,\dots,n$, $l=1,\dots,K_j$ are non-negative multipliers.
\begin{proof}
Optimisation problem $P$ is convex and satisfies the Slater condition, hence strong duality holds. The Lagrangian is 
\small
\begin{align*}
&\mathcal{L}(\tilde{\m{x}},\tilde{\m{s}}, \boldsymbol \pi , \boldsymbol \lambda, \boldsymbol \nu, \boldsymbol \Theta) =  \sum_{f \in F} {\tilde{s}(f)} \\
&+ \sum_{i=1}^n \sum_{f \in F_i} {\lambda_f} \Bigg ( \log \frac{e^{\tilde{x}_i} \sum_{k=1}^{K_i} \pi_{ik} v_{ikf} }{X(e^{\tilde{\m{x}}})}\frac{D_f}{T_s}  - \tilde{s}(f) \Bigg )\\ 
&+ \sum_{i=1}^n \nu_i \Bigg ( 1- \sum_{k=1}^{K_i} \pi_{ik}  \Bigg )  + \sum_{i=1}^n \sum_{k=1}^{K_i} \theta_{ik} \pi_{ik} \notag
\end{align*}
\normalsize
where multipliers $\boldsymbol {\lambda} = [\lambda_{1},\dots,\lambda_{|F|}]^T$, $\boldsymbol {\Theta}= [\boldsymbol \theta_{1},\dots,\boldsymbol \theta_n]^T$ and $\boldsymbol \nu = [\nu_1,\dots,\nu_n]^T$ with $\boldsymbol \theta_i=[\theta_{i1},\dots,\theta_{i|K_i|}]^T$.  
The main KKT conditions are: 
\small
\begin{align}
&\lambda_f = 1 \label{eq:fkkt1},\\
&\sum_{f \in F_j} \lambda_f - \sum_{i=1}^n \Bigg (\frac{x_j}{X(\m{x})}  \prod_{k=1,k \ne j}^{n} (1+x_k) \sum_{f \in F_i} \lambda_f \Bigg )= 0 \label{eq:fkkt2}, \\
&\sum_{f \in F_j} \lambda_f \frac{v_{jlf} }{\sum_{k=1}^{K_j} {\pi_{jk}} v_{jkf} } = \nu_j  - \theta_{jl} \label{eq:fkkt3}.
\end{align}
\normalsize
Claim (i): From the second KKT condition (\ref{eq:fkkt2}), substituting $\lambda_f=1$ and rearranging terms we obtain
\begin{align}
\frac{ |F_j|  }{ {|F| }} =  \frac{x_j }{X(\m{x})} \Bigg (1+ \frac{P_{coll,j}}{1-P_{coll,j}} \Bigg ) =: T_j \label{eq:secondkkt2}
\end{align}
provided $|F|  \ne 0$.  Claim (ii) that $\sum_{i=1}^n T_i=1$ follows  immediately from (\ref{eq:secondkkt2}).   Claim (iii) follows from the third KKT condition (\ref{eq:fkkt3}).   
\end{proof}
\end{theorem}

Note that property (ii), that station airtimes sum to unity,  in Theorem \ref{lem:kkt} extends to MU-MIMO WLANs the result in \cite{DBLP:journals/corr/abs-1206-3120} that this airtime constraint characterises the WLAN rate region boundary.

\subsection{Determining station transmission attempt probability}
Determining the station transmission attempt rates $\m{x}$ requires meeting the constraint that the sum of airtimes sums to unity, and so requires knowledge of all station airtimes in the WLAN.  However, as discussed in \cite{5910091}, decentralised approximations can be found based on local observations of channel idle time.

\subsection{Determining $\nu_j$, $\theta_{jl}$}
The proportional fair rate allocation depends on multipliers $\nu_j$, $\theta_{jl}$.   These depend on the distribution of flows amongst the wireless stations, and on the available MU-MIMO transmission patterns at each station and so cannot be stated in closed-form.   However, they can be readily determined using standard sub-gradient methods.  Namely, by iterating on update $\nu_j(t+1) = \nu_j(t) + \alpha ( 1- \sum_{k=1}^{K_i} \pi_{ik} )$, $\theta_{jl}(t+1)=\theta_{jl}(t)+\alpha \pi_{ik}$, where $\alpha>0$ is a sufficiently small step-size parameter.   Since these updates make use only of information which is locally available at station $j$ they can be implemented in a fully decentralised manner (with no need for message passing).

\subsection{Determining the MU-MIMO transmission pattern}\label{sec:patterns}
The proportional fair transmission pattern conditions (\ref{eq:3kkt}) can be expressed in matrix form as
\begin{align}\label{eq:matrix}
   \m{V}_{j}  (\m{V}_j^T\boldsymbol{\pi}_j)^{-*} = \nu_j \m{1}- \boldsymbol{\theta}_{j},\ j=1,\dots, n
\end{align}
where $\m{x}^{-*}:=[\frac{1}{x_1},\dots,\frac{1}{x_n} ]^T$ for vector $\m{x}=[x_1,\dots,x_n]^T$ and $\m{1}$ denotes the all ones column vector.  

When $\m{V}_{j}$ has full column rank $|F_j|$ (this is commonly satisfied e.g. when the set of possible transmit patterns admits the option to transmit each flow separately in which case $\m{V}_{j}$ contains the $|F_j|\times |F_j|$ identity matrix), then we can write $\m{V}_{j} := \left[\begin{array}{c} \m{X} \\ \m{Y}\end{array} \right]$ where $\m{X}$ is full rank and the rows of $\m{Y}$ are linear combinations of the rows of  $\m{X}$.  This partitioning can always be achieved simply by ordering the rows of $\m{V}_{j}$ appropriately.    Condition (\ref{eq:matrix}) becomes
\begin{align}\label{eq:kkt0}
\left[\begin{array}{c} \m{X} \\ \m{Y}\end{array} \right]  (\m{V}_{j}^T \boldsymbol {\pi}_j)^{-*} = \nu_j \m{1}- \boldsymbol{\theta}_{j}.
\end{align}
Premultiplying both sides by $\left[\begin{array}{cc} \m{X}^{-1}& \m{0}\end{array} \right]$ and re-arranging,
\begin{align}
 \m{V}_{j}^T \boldsymbol {\pi}_j = & \left(\left[\begin{array}{cc} \m{X}^{-1}& \m{0}\end{array} \right]\left(\nu_j \m{1}- \boldsymbol{\theta}_{j}\right)\right)^{-*}. \label{eq:obtainprob}
\end{align}
Given $\nu_j$ and $\boldsymbol{\theta}_{j}$, vectors $\boldsymbol \pi_j$ satisfying (\ref{eq:obtainprob}) can be found using gaussian elimination.   When $\m{V}_{j}$ is non-singular, then the solution to (\ref{eq:obtainprob}) is unique.   However, in general more than one such vector will exist and any such vector is proportional fair.  The RHS of (\ref{eq:obtainprob}) depends only on the multipliers associated with wireless station $j$, which as already noted can be determined using information available locally at station $j$.  Hence, (\ref{eq:obtainprob}) can  be solved to find the proportional fair MU-MIMO transmission patterm for each station $j$ in a fully decentralised manner.

\subsection{Finite load}
Optimisation problem $P$ can be extended to include flow finite offered loads by adding an additional constraint $\tilde{s}_f \le \bar{s}$ for each flow $f$, where $\bar{s}$ is the maximum offered load for flow $f$.  Since these constraints are linear, the optimisation problem remains convex and the foregoing analysis can be directly extended.

\section{Examples}\label{sec:examples}

\subsection{Unequal airtimes with MU-MIMO} \label{ex:unequal}   Consider a WLAN downlink with a MU-MIMO equipped AP that carries 4 flows ($f_1,f_2,f_3,f_4$) transmitted to four client stations.   The offered load is unconstrained.    The matrix of available MU-MIMO transmission patterns at the AP is
\begin{align}
\small
\m{V} = \begin{bmatrix}
     0 & 4  &   0 & 4\\
     2 & 0  &   0 & 1\\
     2 & 2  &   2 & 0\\
     1 & 0  &   4 & 2
\end{bmatrix}.
\label{eq:patterns}
\end{align}
Since the AP is the only transmitter, the optimal $P_{coll,AP}=0$, $\tau_{AP}=1$, and AP total airtime $T_{AP}=1$.   Solving optimisation problem (P), the proportional fair allocation of MU-MIMO transmission patterns is $\boldsymbol \pi =  \left[\frac{1}{3}, 0, \frac{1}{3}, \frac{1}{3} \right]^T$.    

The proportional fair rate allocation often corresponds to an equal airtime allocation.  The appropriate definition to use for flow airtime is not clear when MU-MIMO is used. One option is the total airtime that would be needed by flow $f$ in order to obtain the same throughput when using a single spatial stream, which is given by $T_{i} \sum_{k=1}^{K_i} \pi_{ik}  v_{ikf}$ and is proportional to the average number of spatial streams allocated to the flow.   In the present example, this is $1$ for flow 1 and $2$ for flows 2, 3 and 4.  Another option is the fraction of station $i$ airtime $\sum_{k=1}^{K_i} \pi_{ik}  v_{ikf}/\sum_{f\in F_i} \sum_{k=1}^{K_i} \pi_{ik}  v_{ikf}$ used by flow $f$ spatial streams in this example is 0.1429 for flow 1 and 0.2857 for flows 2, 3 and 4. A third option is the fraction of station transmission opportunities at which a flow transmits, and in this example we have that each flow is scheduled in $2/3$ of the transmissions.   

Importantly, observe that none of these flow airtimes are equal at the proportional fair allocation.   This is because it is the \emph{station} total airtime that corresponds to the shared network resource being consumed and so to the ``cost'' of transmissions.  Indeed this is reflected in Theorem \ref{lem:kkt}.   When MU-MIMO transmission is available, flow transmissions occur in parallel and so multiple flows can share the same station airtime.  For a given station airtime, the proportional fair allocation of spatial streams maximises  the sum of log flow rates, and this need not correspond to allocating the same number of spatial streams or the same fraction of transmission opportunities to flows.

\subsection{IEEE 802.11ac with Rayleigh fading} \label{sec:exampleac} 

We extend the previous example to make the proportional fair allocation of transmission patterns depend on the network characteristics. Consider the WLAN set up of Example \ref{ex:unequal} with an IEEE 802.11ac AP,  channel bandwidth of 20 MHz and guard interval of $800$ ns. For simplicity we assume that all spatial streams use BPSK $1/2$ modulation and coding scheme, and that the transmission power is equally divided amongst the spatial streams in a transmission pattern. We further assume for simplicity that the fading is independent for each antenna in the WLAN and that the AP has full knowledge of the channel. We consider two types of schedulers, proportional fair and uniform, i.e. transmission patterns are allocated the same fraction of transmission opportunities. Regarding the channel we use Rayleigh fading.

\begin{figure}
\centering
\includegraphics[width=.85\textwidth]{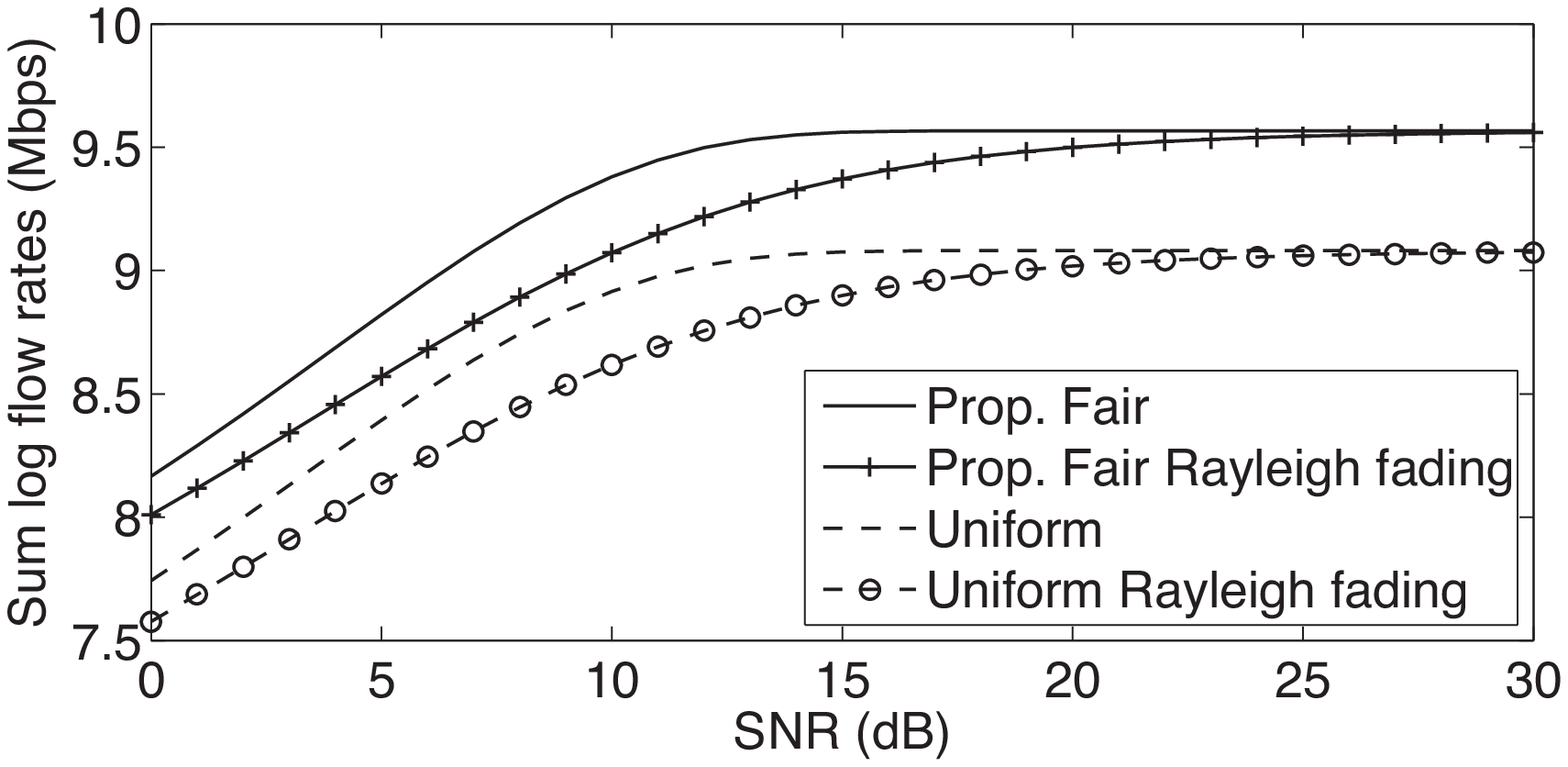}
\caption{Sum of log flow rates in Example \ref{sec:exampleac} for the proportional fair and uniform schedulers with and without Rayleigh fading.}
\label{fig:rate}
\end{figure}

\begin{figure}
\centering
\includegraphics[width=.85\textwidth]{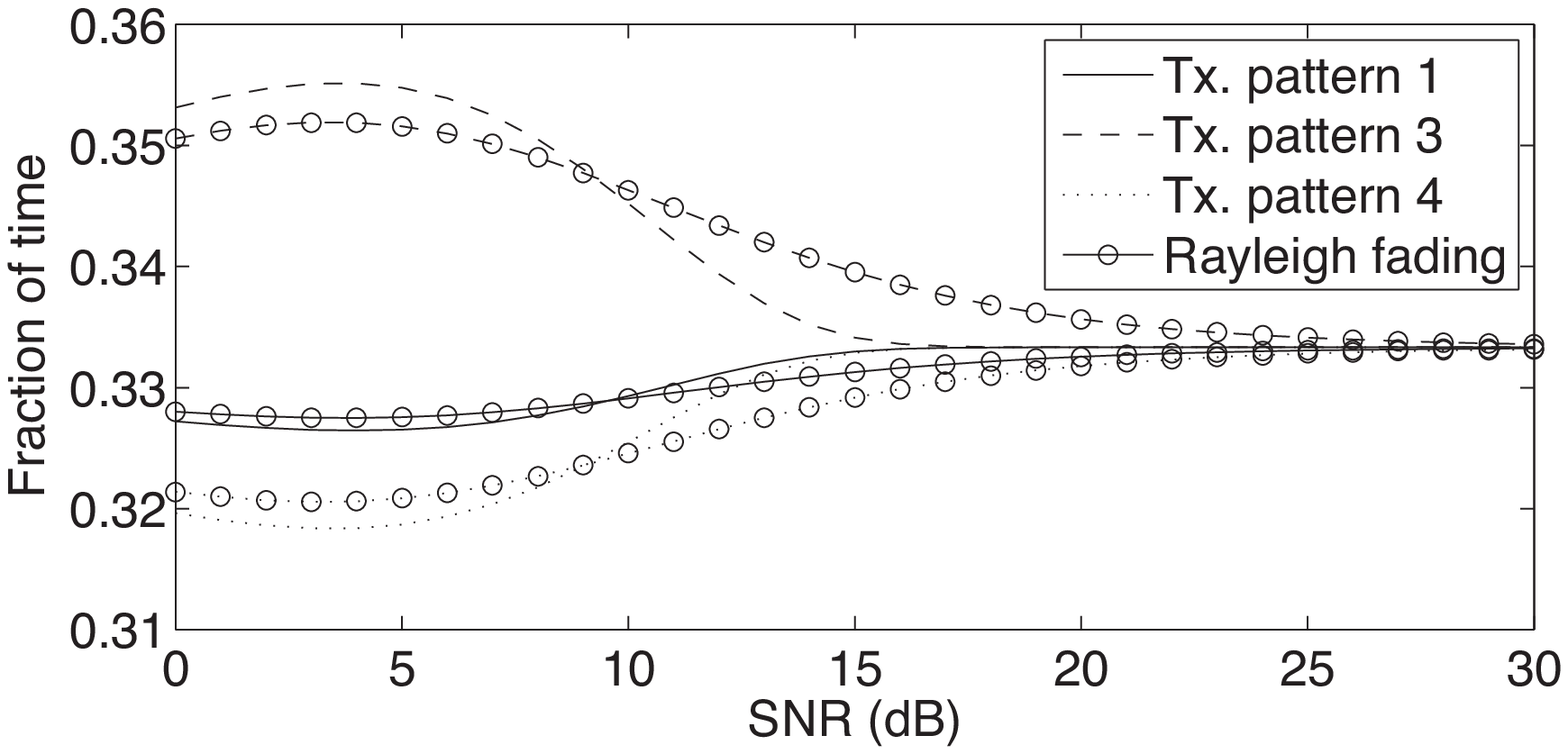}
\caption{Fraction of time that transmission patterns are selected in Example \ref{sec:exampleac} for the proportional fair scheduler with and without Rayleigh fading. Transmission patterns correspond to the rows of matrix $\m{V}$. Transmission pattern 2 is never selected.}
\label{fig:time}
\end{figure}

Notice that differently from Example \ref{ex:unequal}, now in optimisation problem $P$ we have to use the rate as in (\ref{eq:flownonconvex2}) in order to take into account the network characteristics. That is, matrix $\m{D}$ in (\ref{eq:flownonconvex2}) depends on the SNR because it contains the number of bits that can be transmitted for each flow in each transmission pattern. See in Figure \ref{fig:rate} how the sum of log flow rates depends on the SNR, scheduler and fading. Next, observe in Figure \ref{fig:time} how the proportional fair allocation of the transmission patterns, and so the flow's airtimes, changes depending on the SNR and channel characteristics. Moreover, notice that the proportional fair allocation converges to the solution of Example \ref{ex:unequal} when the SNR is large enough. 

\section{Conclusions}
We consider the proportional fair rate allocation in an 802.11 WLAN that supports MU-MIMO transmission by one or more stations.   We characterise, for the first time, the proportional fair allocation of MU-MIMO spatial streams and station transmission opportunities.  While a number of features carry over from the case without MU-MIMO, in general neither flows nor stations need to be allocated equal airtime when MU-MIMO is available.

\bibliographystyle{unsrt}
\bibliography{valls_WCL_IEEEtran.bib}

\end{document}